## Discovery and outburst characteristics of the dwarf nova ASAS J224349+0809.5

Jeremy Shears, Patrick Wils, Greg Bolt, Franz-Josef Hambsch, Tom Krajci, Ian Miller, Richard Sabo and Bart Staels

### Abstract

We report the discovery of a new dwarf nova and our observations of its first confirmed superoutburst during 2009 October. The outburst amplitude was 6 magnitudes. The main outburst lasted 17 days and was followed 4 days later by a remarkable rebrightening. Superhumps were present during the main outburst, which confirms that it belongs to the SU UMa family. Initially the mean superhump period was $P_{sh}$ = 0.06965(17) d, but analysis of the O-C residuals showed a dramatic evolution in $P_{sh}$ during the outburst. During the first two-thirds of the plateau phase the period increased with $dP_{sh}/dt = +1.24(5) \times 10^{-3}$. There was then an abrupt change following which the period decreased with $dP_{sh}/dt = -1.01(9) \times 10^{-3}$. The amplitude of the superhumps also varied, with a maximum amplitude near the beginning of the outburst and a second maximum corresponding to the discontinuity in $P_{sh}$. Analysis of archival data showed outbursts also occurred in October 2005, June 2006 and June 2007. Assuming that the superoutbursts are periodic, we estimate that the outburst period is around 450 days.

### Introduction

Dwarf novae are a class of cataclysmic variable star in which a white dwarf primary accretes material from a secondary star via Roche lobe overflow. The secondary is usually a late-type main-sequence star. In the absence of a significant white dwarf magnetic field, material from the secondary passes through an accretion disc before settling on the surface of the white dwarf. As material builds up in the disc, a thermal instability is triggered that drives the disc into a hotter, brighter state causing an outburst in which the star apparently brightens by several magnitudes [1]. Dwarf novae of the SU UMa family occasionally exhibit superoutbursts which last several times longer than normal outbursts and may be up to a magnitude brighter. During a superoutburst the light curve of a SU UMa system is characterised by superhumps. These are modulations in the light curve which are a few percent longer than the orbital period. They are thought to arise from the interaction of the secondary star orbit with a slowly precessing eccentric accretion disc. The eccentricity of the disc arises because a 3:1 resonance occurs between the secondary star orbit and the motion of matter in the outer accretion disc. For a more detailed review of SU UMa dwarf novae and superhumps, the reader is directed to reference 1.

In this paper we present our discovery of a new dwarf nova, photometry of its first confirmed superoutburst and analysis of archival outbursts.

## Discovery

The object was found in the course of a systematic search for dwarf novae in archival images which is currently being undertaken by one of the authors (PW). Full details of the techniques and results will appear elsewhere [2], but the basic method is as follows. A selection of blue objects was made from the GALEX survey Data Release 3 [3]. These were then matched to objects in the USNO-B1.0 catalogue [4] via VizieR [5]. Those objects showing a large magnitude variation in the USNO-B1.0 catalogue were then selected and checked on archival images from the United States Naval Observatory Flagstaff Station [6], which include the images from which the USNO-B1.0 catalogue was originally created, to confirm whether they were real objects or artefacts. One of these objects was USNO-B1.0 0981-0723427, which was found to be in outburst on one of the near-infrared images from 1995, but faint on another nine images (Figure 1). Further outbursts were identified in ASAS-3 data [7]. These characteristics are strongly suggestive that the object is a dwarf nova and an announcement of the likely dwarf nova identity of ASAS J224349+0809.5 was made on the CVnet-discussion group [8]. Subsequently, PW identified a further outburst in ASAS-3 data on 2009 Oct 6 at V=12.8 and announced it on the CVnet-outburst group [9], allowing intensive CCD photometry to be conducted, the results of which are discussed in this paper.

## Photometry and analysis

The authors conducted V-filtered and unfiltered photometry using the instrumentation shown in Table 1 and according to the observation log in Table 2. Images were dark-subtracted and flat-fielded prior to being measured using differential aperture photometry relative to one or more of the following stars: GSC 1152-0973 (V=13.06), 1152-1047 (V=11.75), 1152-1059 (V=12.89); magnitudes from GSC 2.2. Heliocentric corrections were applied to all data.

## Profile of the October 2009 outburst

The overall light curve of the outburst is shown in the top panel of Figure 2, based on the authors' photometry, supplemented with data from the Catalina Real-Time Sky Survey [10] (CRTS) and data from Eddy Muyllaert supplied via the AAVSO International Database. The star was observed to be at its brightest at V=12.8 in the ASAS-3 discovery image. CRTS data show that the star varies been V= 18.2 and 19.1 in quiescence with a mean of 18.7. Thus the outburst amplitude is about 6 magnitudes above mean quiescence.

Time series photometry commenced on JD 2455113, some 3 days after the outburst was first detected by ASAS-3. There is some evidence that the star was brightening during JD 2445113, which may indicate that the ASAS-3 detection was actually a precursor outburst. The period from JD 2445113 and 2455124 corresponds to the plateau phase during which the star faded gradually at a mean rate of 0.11 mag/d. There then followed a rapid decline (>0.98 mag/d) over the next 3 days reaching mag ~17.8 some 17 days after the outburst was detected. This temporary minimum lasted for 4 days, following which it rebrightened to mag 14.0. The duration of this second brightening event was about 3 days, after which it returned to quiescence at 18.7. Including the second brightening, the outburst lasted about 26 days.

**Measurement of the superhump period**

We plot expanded views of the longer time series photometry runs, having subtracted the mean, in Figure 3 where each panel shows 2 days of data drawn to the same scale. This clearly shows the presence of regular modulations which we interpret as superhumps. The presence of superhumps is diagnostic that ASAS J224349+0809.5 is a member of the SU UMa family of dwarf novae, making this the first confirmed superoutburst of the star.

To study the superhump behaviour, we first extracted the times of each sufficiently well-defined superhump maximum by fitting a quadratic function to the individual light curves around each maximum. Times of 63 superhump maxima were found and are listed in Table 3. An analysis of the times of maximum for cycles 0 to 54 (JD 2455113 to 2455116) allowed us to obtain the following linear superhump maximum ephemeris:

$$HJD_{max} = 2455113.04223(31) + 0.06965(17) \times E \qquad \text{Equation 1}$$

This gives the mean superhump period for the first three days of the superoutburst $P_{sh}$ = 0.06965(17) d. The O–C residuals for the superhump maxima for the complete outburst relative to the ephemeris are shown in the middle panel in Figure 2.

**Superhump evolution**

The O-C diagram in Figure 2 (middle panel) shows that the superhump period changed significantly during the outburst. Kato *et al*. [11] studied the superhump period evolution in a large number of SU UMa systems and found that many outbursts appeared to show three distinct stages: an early evolutionary stage (A) with longer superhump period, a middle stage (B) during which systems with $P_{orb} <$ 0.08 d have a positive period derivative, and a final stage (C) with a shorter $P_{sh}$. Our O-C diagram for ASAS J224349+0809.5 is very similar to O-C diagrams of SW UMa and UV Per, amongst many others which exhibit this evolutionary trend, although it appears we missed stage A. The interval between JD 2455113 and 2455119 (red data points in Figure 2) corresponds to Stage B, during which we found an increase in the superhump period with $dP_{sh}/dt = +1.24(5) \times 10^{-3}$ by fitting a quadratic function to the data (red data points and dotted red line in Figure 2). There was then a change in period at JD 2455120, following which the period decreased with $dP_{sh}/dt = -1.01(9) \times 10^{-3}$ during stage C as shown by the quadratic fit to the data between JD 2455120 and JD 2455128 (green data points and trend line in Figure 2). The period transition from stage B to C was abrupt and the period change discontinuous, as is common in SU UMa systems [11].

We also found that the superhump peak-to-peak amplitude changed significantly during the outburst and had two maxima (Figure 2, bottom panel). During the first night of observation (JD 2455113), the amplitude increased from 0.17 to 0.21 magnitude, after which it gradually decreased to 0.11 magnitude on JD 2455117. The superhumps then increased in amplitude for a second time, reaching a second maximum of 0.23 magnitudes on JD 2455120, after which they again declined to 0.11 magnitudes. The second maximum corresponds to the superhump period change discontinuity between stage B and C. A similar bimodality in superhump maxima was also observed in SW UMa [11, 12] where the second peak also coincided with the transition from stage B to C. The superhumps may have grown again during the temporary minimum (see the final panel in Figure 3), but there was considerable scatter in the data (noise in the photometry as a result of the faintness of the object) which made it difficult to measure their amplitude.

**Orbital period**

Gaensicke et al. [13] developed an empirical relation between $P_{orb}$ and $P_{sh}$ by analysing a wide range of SU UMa systems:

$$P_{orb} = 0.9192(52)\ P_{sh} + 5.39(52) \qquad \text{Equation 2}$$

where both periods are given in minutes. Using our value of $P_{sh}$ = 0.06965(17) d in this equation, allows us to estimate $P_{orb}$ = 0.06777(88) d.

**Outburst frequency**

Southworth *et al.* [14] presented a method to estimate the length of the superoutburst cycle of an SU UMa system, based on the typical superoutburst duration, the observed number of outbursts and the actual times of observation. In the case of ASAS J224349+0809.5, the outburst light curve shows that the object was brighter than mag 14.5 (representing a magnitude at which it could be picked up by ASAS-3) for about 15 days and the total outburst duration, including the temporary minimum, was about 18 days (representative of the length of time during which CRTS could detect the outburst, it having a fainter detection limit than ASAS-3). We do not include the rebrightening event in the latter estimate as we have no way of knowing if it is an unusual phenomenon in this system. Monte Carlo simulations then show that there is a 45% probability of ASAS-3 observing an outburst and 47% for CRTS. Combining the data sets gives a probability to observe an outburst of a little above 50%. Both ASAS-3 (using A or B quality data only) and CRTS have observed four outbursts (October 2005, June 2006, June 2007, October 2009). Assuming that the superoutbursts are quasi-periodic (i.e. with the length of the superoutburst cycle following a Gaussian distribution, with a standard deviation of a quarter of its mean), further Monte Carlo simulations suggest a period between 250 and 900 days, with a period of around 450 days as the most likely.  Figure 4 shows a plot of the probability of observing 4 outbursts for a given length of superoutburst period using the CRTS and ASAS times of observation.

**Discussion**

One of the most intriguing aspects of the superoutburst of ASAS J224349+0809.5 is the post-outburst rebrightening. We suspect that this was actually part of the outburst rather than a separate "normal" outburst, because (i) the star was about a magnitude above true quiescence between the main outburst and the rebrightening (ii) superhumps were still present during this temporary minimum, showing that the accretion disc was still eccentric and (iii) it occurred so soon after the main outburst. Similar rebrightening events immediately following a superoutburst have been reported in several SU UMa systems including GO Com [15], V725 Aql [16], GSC2.3 N152008120 [17] and V1028 Cyg [18]. Post-outburst rebrightenings are most commonly observed in a sub-group of SU UMa systems, the WZ Sge stars, where even multiple rebrightenings are often observed. For example, following the 1996 outburst of EG Cnc, six consecutive rebrightenings were observed with a mean interval of about seven days before it dropped back to full quiescence [19]. WZ Sge stars are highly evolved dwarf novae having long intervals between outbursts, typically decades, and exceptionally large outburst amplitudes, typically 6 to 8 magnitudes, and very short orbital periods. Osaki et al. [20, 21] have modelled the rebrightening phenomenon and considered why it occurs mostly in WZ Sge stars. They propose that large amounts of material accumulates in the accretion disc during the long quiescence and this in turn leads to their large amplitude and long duration outbursts. The quantity of material is such that during the superoutburst the disc extends beyond the 3:1 resonance radius which exists between the secondary star orbit and the motion of matter in the outer accretion disc surrounding the white dwarf primary. At the end of a superoutburst a cooling wave propagates inwards, but if some regions of the disc are left close to the critical surface density required for an outburst, a small amount of fresh material fed into these regions can cause a rebrightening. Normally, superhumps are still present during rebrightening episodes since the disc still extends beyond the 3:1 resonance radius. Unfortunately our data during the rebrightening were not of sufficient quality to be certain whether or not superhumps were present.

We note that the SU UMa systems mentioned above which have shown post-outburst rebrightenings, GO Com, V725 Aql, GSC2.3 N152008120, V1028 Cyg plus ASAS J224349+0809.5 itself, all had large outburst amplitudes of the order of 6 magnitudes. Such a large amplitude suggests that large amounts of material had accumulated in the accretion disc prior to outburst, so, following Osaki *et al*.'s [20, 21] model, one can speculate that the disc was also extended beyond the 3:1 resonance radius and the rebrightening is phenomenologically similar to the situation in WZ Sge systems, i.e. the extension makes it more likely that there will be sufficient material left over when the cooling wave propagates to reignite the outburst.

This leads us to consider whether ASAS J224349+0809.5 could itself be a WZ Sge system. Firstly, its superoutburst amplitude (~6 magnitudes) is similar to WZ Sge systems, although it is still within the range of normal SU UMa systems. Secondly, WZ Sge systems have short orbital periods and they become more common at the lower end of the orbital distribution of SU UMa systems, towards the period minimum (~78 min or 0.054 d [1]). However, the likely orbital period of ASAS J224349+0809.5 is rather longer than the majority of WZ Sge systems, although it is shorter than RZ Leo ($P_{orb}$ = 0.07651 d [22]), which has the longest of WZ Sge systems [23]. And thirdly, the outburst period of about 450 days for ASAS J224349+0809.5 is much less than for most WZ Sge systems and more typically of normal SU UMa systems. This leads us to conclude that it is most unlikely that ASAS J224349+0809.5 is a WZ Sge and is more likely a typical SU UMa system. However, spectroscopic evidence is required to be certain.

Our data also suggest that ASAS J224349+0809.5 underwent a precursor outburst before the main superoutburst. From the published superoutburst light curves, it appears that precursor outbursts occur in some, but by no means all SU UMa dwarf novae, and no global study on the presence or absence of such events has been made so far. We note that superhumps were already present during the rise to superoutburst (JD 2455113). A similar situation occurred during the 2008 superoutburst of HS 0417+7445 (= 1RXS J042332+745300) where a precursor outburst appeared to trigger the superoutburst and superhumps were observed during the rise to superoutburst maximum [24]. By contrast in the case of GO Com, a precursor was seen in the 2003 superoutburst, although superhumps did not appear to be present during this stage [25]. Clearly there is value in documenting the presence and behaviour of precursor outbursts in other SU UMa systems.

## Conclusions

We report the discovery a new dwarf nova, ASAS J224349+0809.5. We subsequently observed superhumps during the first confirmed superoutburst of this star in 2009 October, which shows that is a member of the SU UMa family. The outburst amplitude was 6 magnitudes. The first 11 days of our time series photometry corresponded to the plateau phase, after which the star faded rapidly to a temporary minimum, some 17 days after the detection of the outburst. Remarkably, some 4 days later the star rebrightened for about 3 days, after which it returned to quiescence. Including the rebrightening, the outburst lasted some 26 days.

We determined the mean superhump period from our first 3 days of observations as $P_{sh}$ = 0.06965(17) d, however analysis of the O-C residuals showed a dramatic evolution in $P_{sh}$ during the outburst. During the first two-thirds of the plateau phase the period increased with $dP_{sh}/dt = +1.24(5) \times 10^{-3}$. There was then an abrupt change following which the period decreased with $dP_{sh}/dt = -1.01(9) \times 10^{-3}$. The amplitude of the superhumps also varied, with a maximum amplitude near the beginning of the outburst and a second maximum corresponding to the discontinuity in $P_{sh}$.

We estimate the orbital period of ASAS J224349+0809.5 as $P_{orb}$ = 0.06777(88) d, based on the measured superhump period.

Analysis of archival data showed outbursts had also occurred in October 2005, June 2006 and June 2007. Assuming that the superoutbursts are periodic, we estimate that the outburst period is around 450 days.

We propose that ASAS J224349+0809.5 is likely to be a normal SU UMa system, rather than a more highly evolved WZ Sge system.

## Acknowledgements

The authors gratefully acknowledge observations made by Eddy Muyllaert which we obtained via the AAVSO International Database, the use of SIMBAD and Vizier, operated through the Centre de Donées Astronomiques (Strasbourg, France), and the NASA/Smithsonian Astrophysics Data System. JS thanks the Council of the British Astronomical Association for the award of a Ridley Grant that was used to purchase some of the equipment that was used in this research. WE thanks both referees, Drs. Tim Naylor and Chris Lloyd for help comments that have improved the paper

## Addresses

JS: “Pemberton”, School Lane, Bunbury, Tarporley, Cheshire, CW6 9NR, UK [bunburyobservatory@hotmail.com]
PW: Vereniging voor Sterrenkunde, Belgium [patrickwils@yahoo.com]
GB: CBA Perth, 295 Camberwarra Drive, Craigie, Western Australia 6025, Australia [gbolt@iinet.net.au]
FJH: Vereniging voor Sterrenkunde, CBA Mol, Belgium, AAVSO, GEOS, BAV [hambsch@telenet.be]
TK: CBA New Mexico, PO Box 1351 Cloudcroft, New Mexico 88317, USA [tom_krajci@tularosa.net]
IM: Furzehill House, Ilston, Swansea, SA2 7LE, UK [furzehillobservatory@hotmail.com]
RS: 2336 Trailcrest Dr., Bozeman, MT 59718, USA [richard@theglobal.net]
BS: CBA Flanders, Patrick Mergan Observatory, Koningshofbaan 51, Hofstade, Aalst, Belgium [staels.bart.bvba@pandora.be]

| Observer | Telescope | CCD |
|---|---|---|
| Bolt | 0.25 m SCT | SBIG ST-7 |
| Hambsch | 0.4 m reflector | SBIG STL-11kXM |
| Astrokolkhoz collaboration team (Hambsch and Krajci) | 0.3 m SCT | SBIG ST9-XME |
| Miller | 0.35 m SCT | Starlight Xpress SXVF-H16 |
| Sabo | 0.43 m reflector | SBIG STL-1001 |
| Shears | 0.28 m SCT | Starlight Xpress SXVF-H9 |
| Staels | 0.28 m SCT | Starlight Xpress MX-716 |

**Table 1: Equipment used**

| Start time | Duration (h) | Filter | Observer |
|---|---|---|---|
| 2455113.010 | 5.54 | C | Bolt |
| 2455113.365 | 4.44 | V | Miller |
| 2455113.677 | 4.27 | V | Hambsch/Krajci |
| 2455114.573 | 6.70 | V | Hambsch/Krajci |
| 2455115.661 | 0.82 | V | Sabo |
| 2455116.296 | 3.89 | V | Miller |
| 2455116.557 | 6.96 | V | Hambsch/Krajci |
| 2455116.674 | 3.26 | V | Sabo |
| 2455117.288 | 4.10 | C | Staels |
| 2455117.303 | 4.85 | V | Miller |
| 2455118.253 | 3.02 | V | Hambsch |
| 2455119.379 | 2.86 | V | Hambsch |
| 2455119.557 | 6.77 | V | Hambsch/Krajci |
| 2455120.378 | 2.14 | V | Hambsch |
| 2455120.553 | 6.79 | V | Hambsch/Krajci |
| 2455121.259 | 5.57 | C | Shears |
| 2455121.552 | 6.74 | V | Hambsch/Krajci |
| 2455122.104 | 2.50 | C | Bolt |
| 2455122.285 | 5.11 | V | Hambsch |
| 2455122.552 | 6.67 | V | Hambsch/Krajci |
| 2455122.629 | 5.02 | V | Sabo |
| 2455123.551 | 2.81 | V | Hambsch/Krajci |
| 2455124.554 | 2.93 | V | Hambsch/Krajci |
| 2455127.558 | 3.89 | V | Hambsch/Krajci |
| 2455127.590 | 3.26 | V | Hambsch/Krajci |
| 2455128.551 | 4.15 | V | Hambsch/Krajci |
| 2455131.651 | 1.66 | V | Hambsch/Krajci |
| 2455133.239 | 6.02 | C | Shears |
| 2455133.241 | 1.78 | C | Hambsch |

**Table 2 : Log of time-series photometry**

| Superhump cycle | Superhump maximum (HJD) | O-C ( d) | Uncertainty (d) |
|---|---|---|---|
| 0 | 2455113.0447 | 0.0025 | 0.0003 |
| 1 | 2455113.1134 | 0.0015 | 0.0004 |
| 2 | 2455113.1842 | 0.0027 | 0.0002 |
| 5 | 2455113.3912 | 0.0007 | 0.0007 |
| 6 | 2455113.4592 | -0.0010 | 0.0004 |
| 7 | 2455113.5308 | 0.0011 | 0.0005 |
| 10 | 2455113.7376 | -0.0011 | 0.0005 |
| 11 | 2455113.8069 | -0.0015 | 0.0003 |
| 23 | 2455114.6425 | -0.0017 | 0.0006 |
| 24 | 2455114.7106 | -0.0032 | 0.0005 |
| 25 | 2455114.7810 | -0.0024 | 0.0005 |
| 26 | 2455114.8509 | -0.0022 | 0.0008 |
| 38 | 2455115.6878 | -0.0011 | 0.0005 |
| 47 | 2455116.3162 | 0.0004 | 0.0003 |
| 48 | 2455116.3870 | 0.0016 | 0.0008 |
| 49 | 2455116.4547 | -0.0004 | 0.0008 |
| 51 | 2455116.5949 | 0.0005 | 0.0004 |
| 52 | 2455116.6670 | 0.0029 | 0.0004 |
| 53 | 2455116.7342 | 0.0005 | 0.0013 |
| 53 | 2455116.7329 | -0.0008 | 0.0011 |
| 54 | 2455116.8042 | 0.0009 | 0.0009 |
| 54 | 2455116.8042 | 0.0009 | 0.0015 |
| 62 | 2455117.3612 | 0.0007 | 0.0003 |
| 63 | 2455117.4362 | 0.0060 | 0.0009 |
| 64 | 2455117.5030 | 0.0032 | 0.0000 |
| 76 | 2455118.3419 | 0.0063 | 0.0018 |
| 91 | 2455119.3919 | 0.0115 | 0.0015 |
| 92 | 2455119.4618 | 0.0118 | 0.0024 |
| 94 | 2455119.6022 | 0.0129 | 0.0003 |
| 95 | 2455119.6728 | 0.0138 | 0.0005 |
| 96 | 2455119.7423 | 0.0136 | 0.0007 |
| 97 | 2455119.8118 | 0.0135 | 0.0003 |
| 106 | 2455120.4373 | 0.0122 | 0.0009 |
| 108 | 2455120.5770 | 0.0126 | 0.0026 |
| 109 | 2455120.6471 | 0.0130 | 0.0005 |
| 110 | 2455120.7163 | 0.0126 | 0.0004 |
| 111 | 2455120.7845 | 0.0112 | 0.0005 |
| 118 | 2455121.2737 | 0.0128 | 0.0026 |
| 119 | 2455121.3402 | 0.0096 | 0.0005 |
| 120 | 2455121.4087 | 0.0085 | 0.0003 |
| 121 | 2455121.4784 | 0.0086 | 0.0003 |
| 123 | 2455121.6186 | 0.0094 | 0.0006 |
| 124 | 2455121.6881 | 0.0093 | 0.0004 |
| 125 | 2455121.7574 | 0.0090 | 0.0005 |
| 126 | 2455121.8256 | 0.0074 | 0.0004 |
| 133 | 2455122.3139 | 0.0082 | 0.0014 |
| 134 | 2455122.3817 | 0.0064 | 0.0024 |
| 135 | 2455122.4513 | 0.0063 | 0.0015 |
| 137 | 2455122.5927 | 0.0085 | 0.0008 |
| 138 | 2455122.6595 | 0.0056 | 0.0024 |
| 140 | 2455122.7971 | 0.0039 | 0.0023 |
| 151 | 2455123.5629 | 0.0036 | 0.0028 |
| 152 | 2455123.6322 | 0.0032 | 0.0020 |
| 166 | 2455124.5998 | -0.0043 | 0.0024 |
| 167 | 2455124.6684 | -0.0054 | 0.0029 |
| 209 | 2455127.5741 | -0.0249 | 0.0033 |
| 210 | 2455127.6395 | -0.0293 | 0.0057 |
| 210 | 2455127.6458 | -0.0229 | 0.0128 |
| 211 | 2455127.7105 | -0.0279 | 0.0090 |
| 211 | 2455127.7113 | -0.0271 | 0.0103 |
| 224 | 2455128.6114 | -0.0325 | 0.0095 |
| 225 | 2455128.6746 | -0.0389 | 0.0089 |
| 268 | 2455131.6677 | -0.0407 | 0.0158 |

**Table 3: Superhump maximum times**

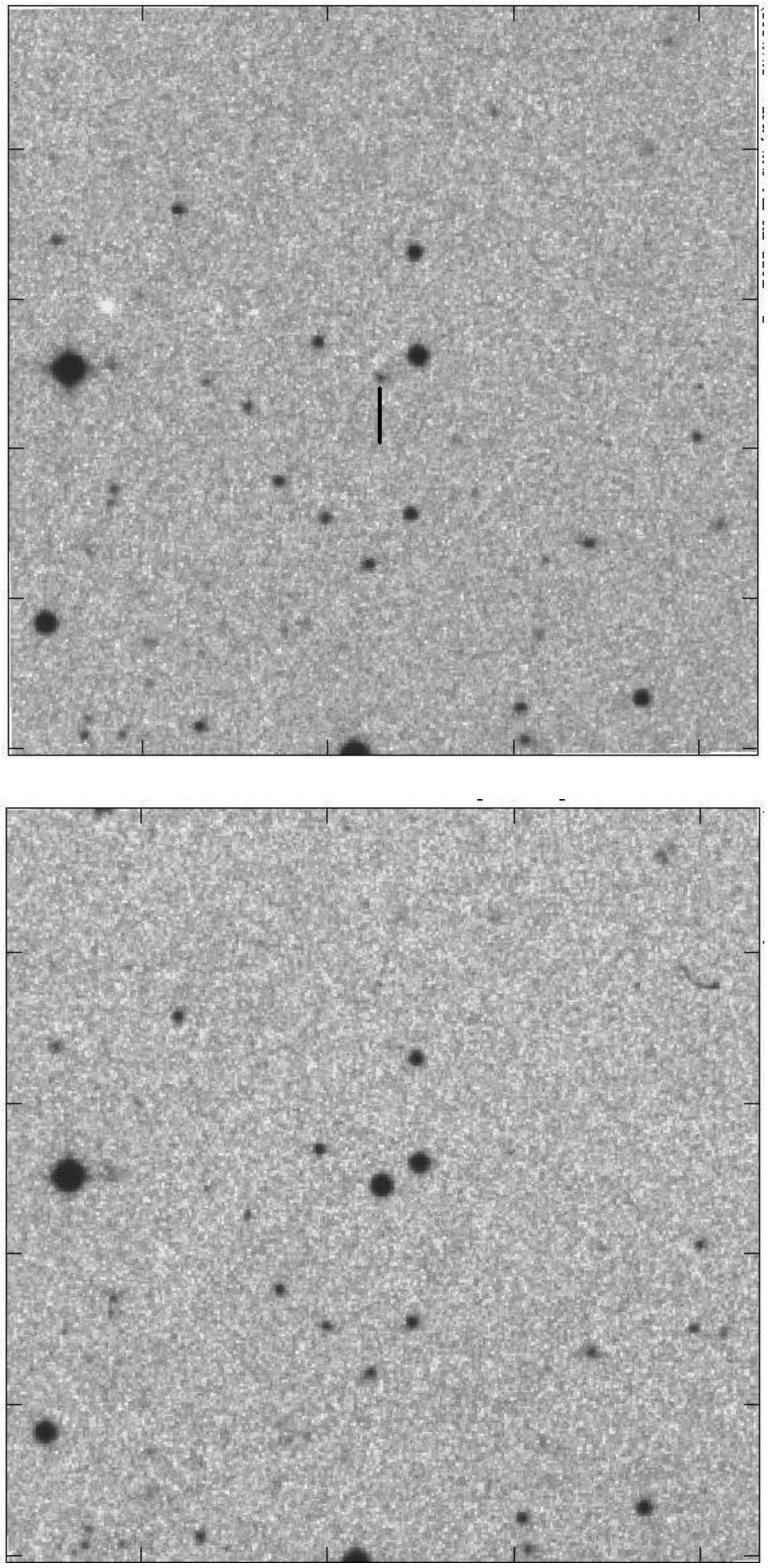

(a)

(b)

**Figure 1: ASAS J224349+0809.5 (a) in quiescence on 1995.5524 (indicated by the bar) and (b) in outburst on 1995.6591**

Images from United States Naval Observatory

Field 5 x 5 min, N at top, E to left

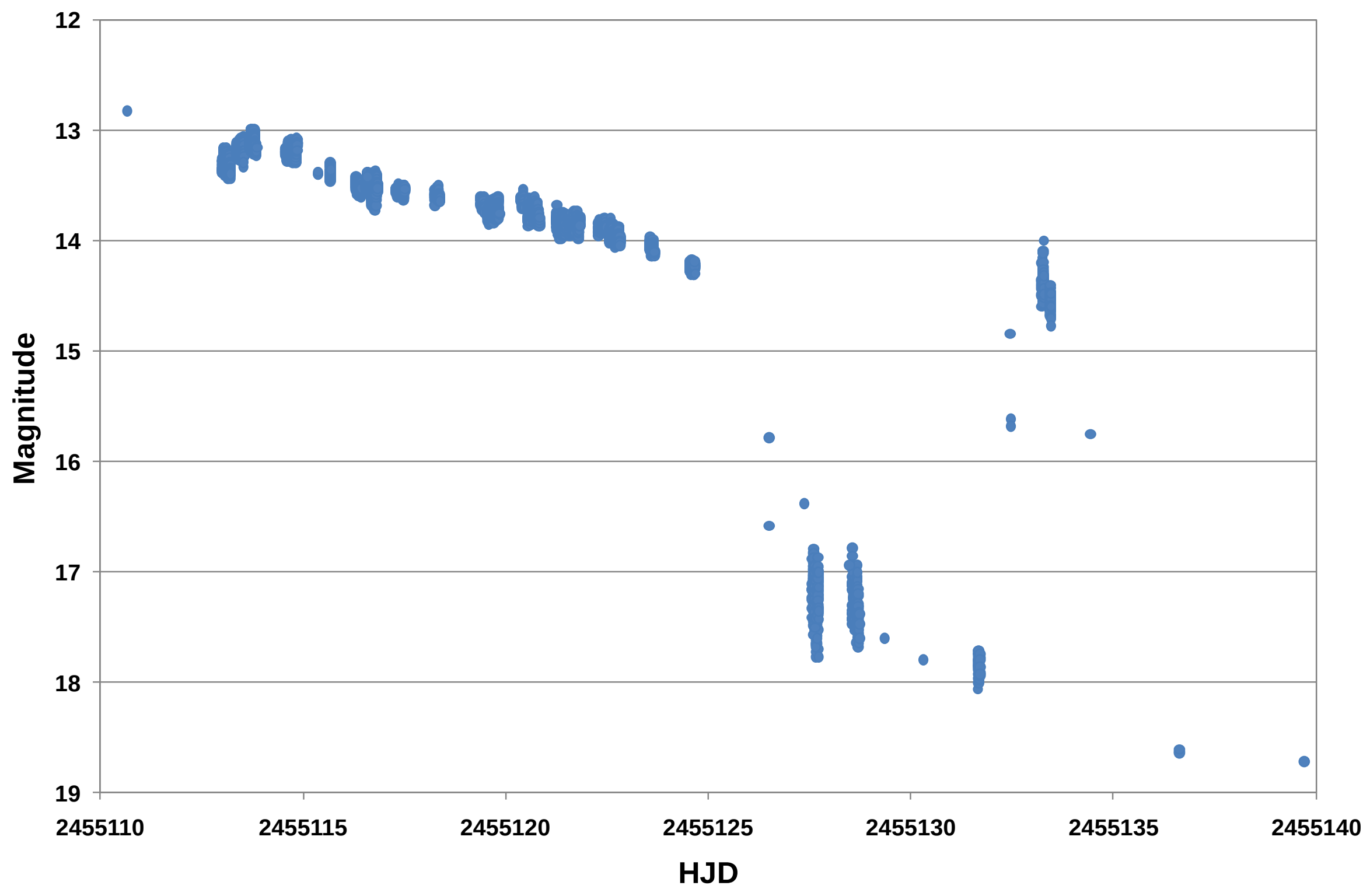

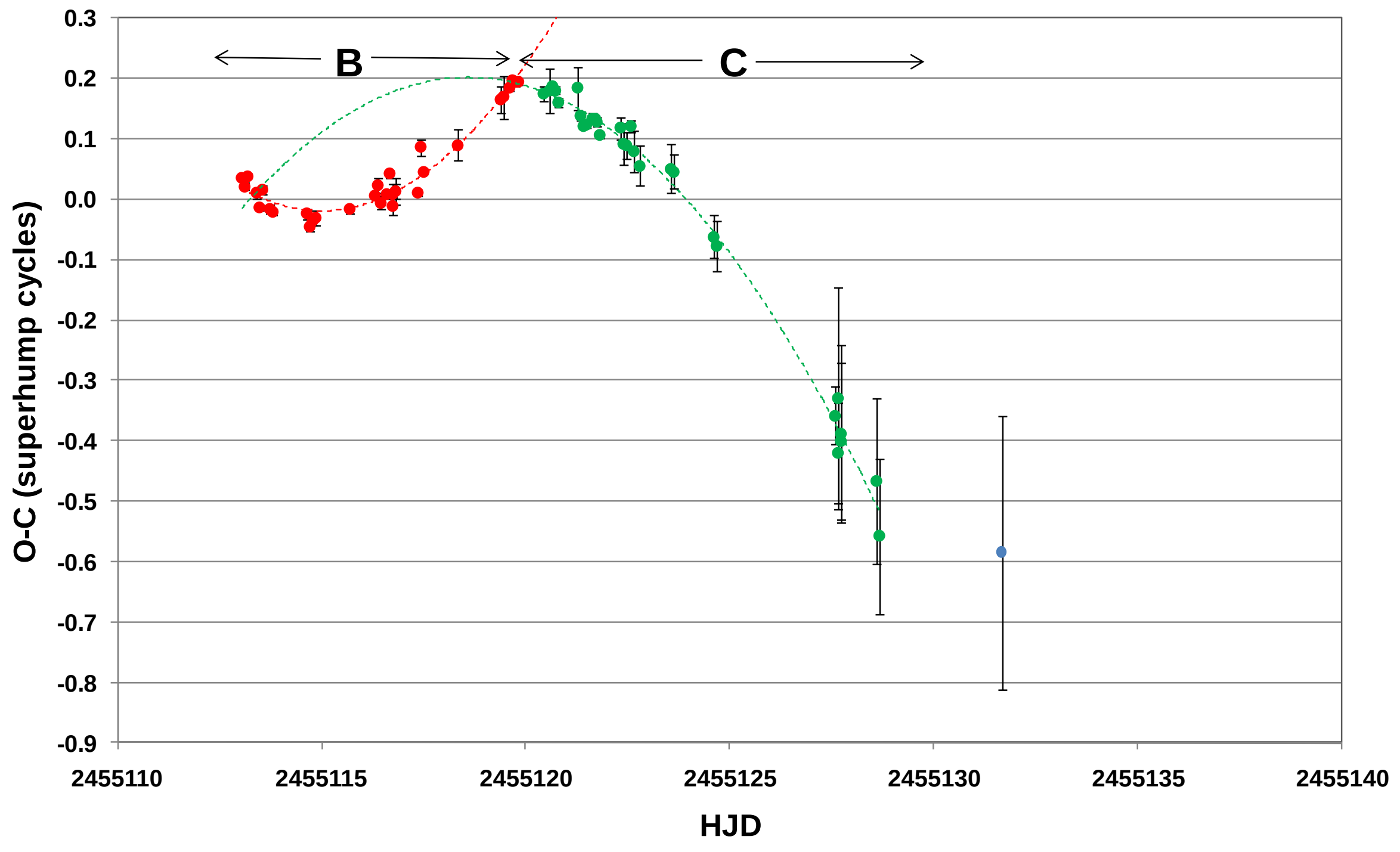

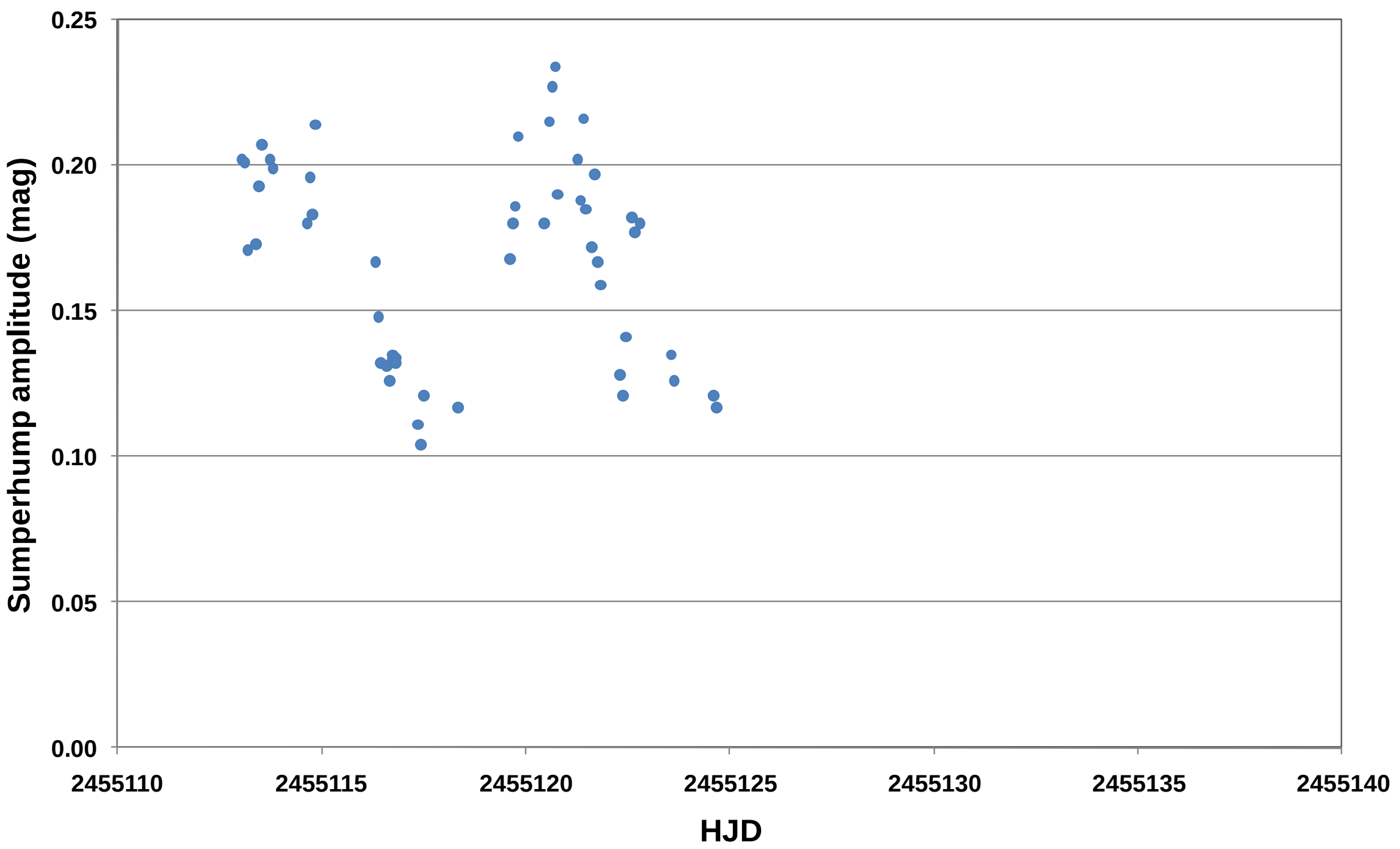

**Figure 2: Light curve of the outburst (top), O-C diagram of superhump maxima relative to the ephemeris in Equation 1 (middle) and superhump amplitude (bottom)**

In the O-C diagram, the red dotted line is a quadratic fit to the data between JD 2455113 and 2455119 (red data points) and the green dotted line is a quadratic fit to the data between JD 2445120 to 2445128 (green data points). The O-C values, presented here in superhump cycles, were obtained by taking the O-C values in Table 3, in days, and dividing by the measured value of $P_{sh}$ = 0.06965 d

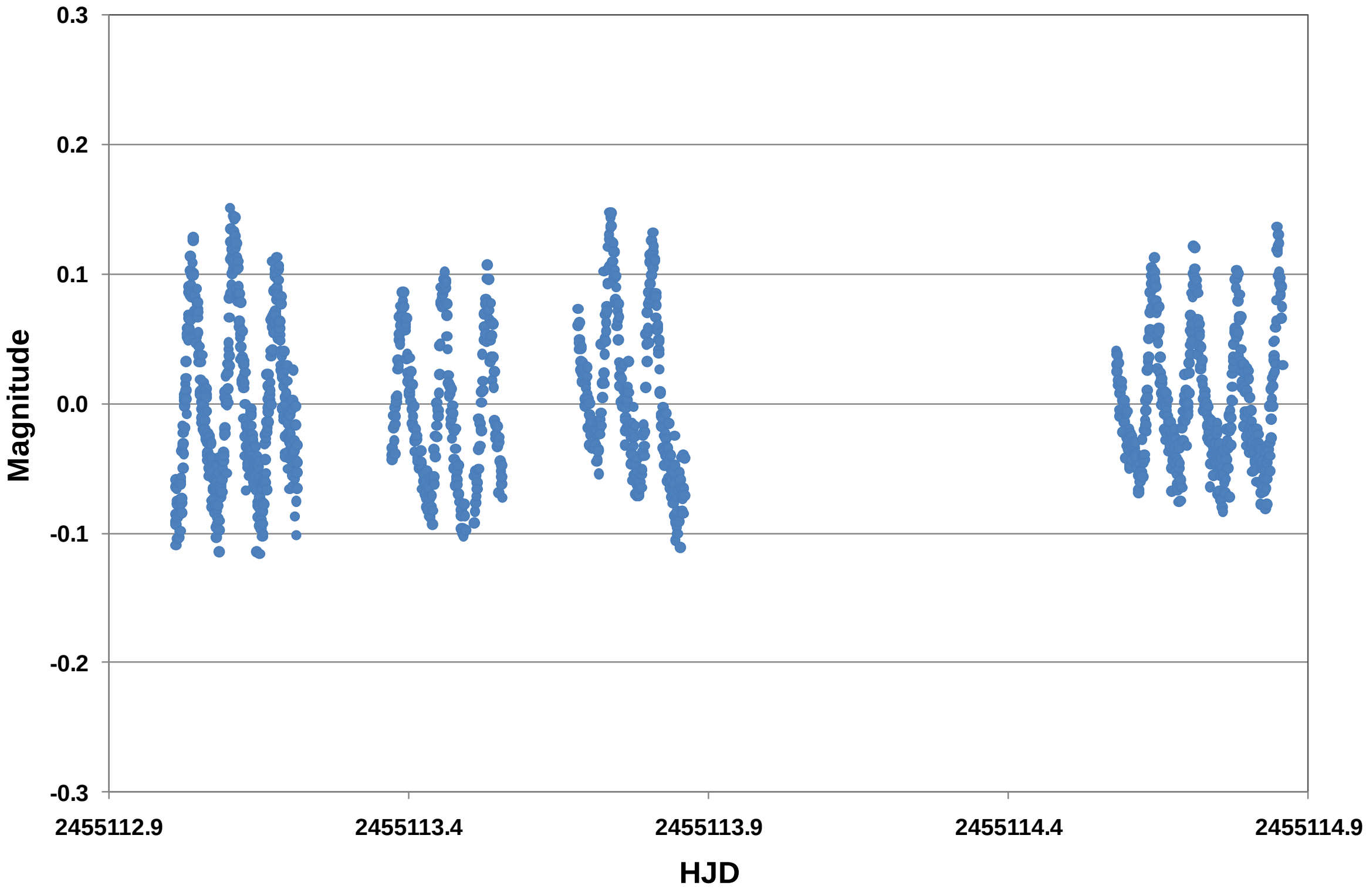

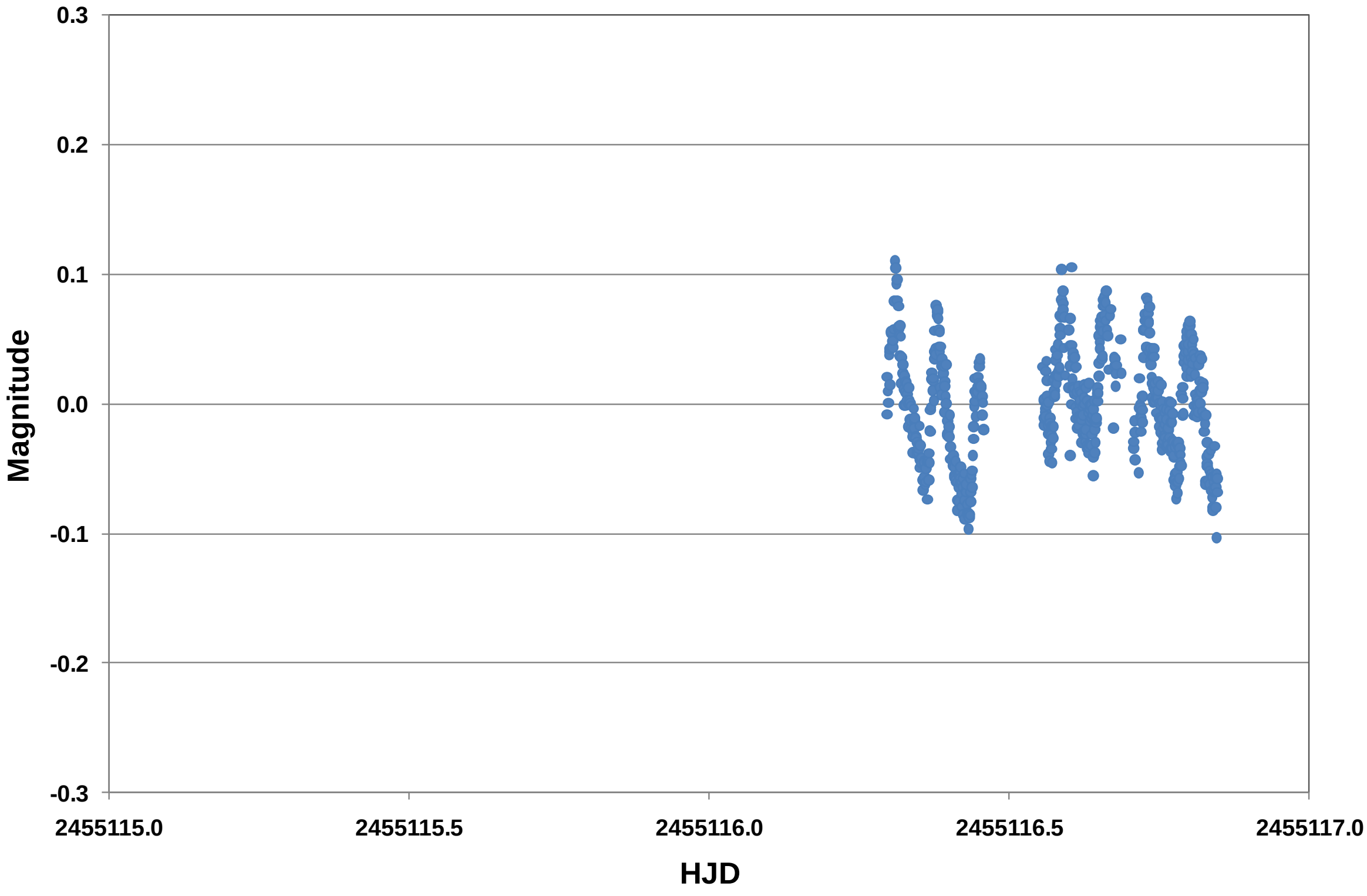

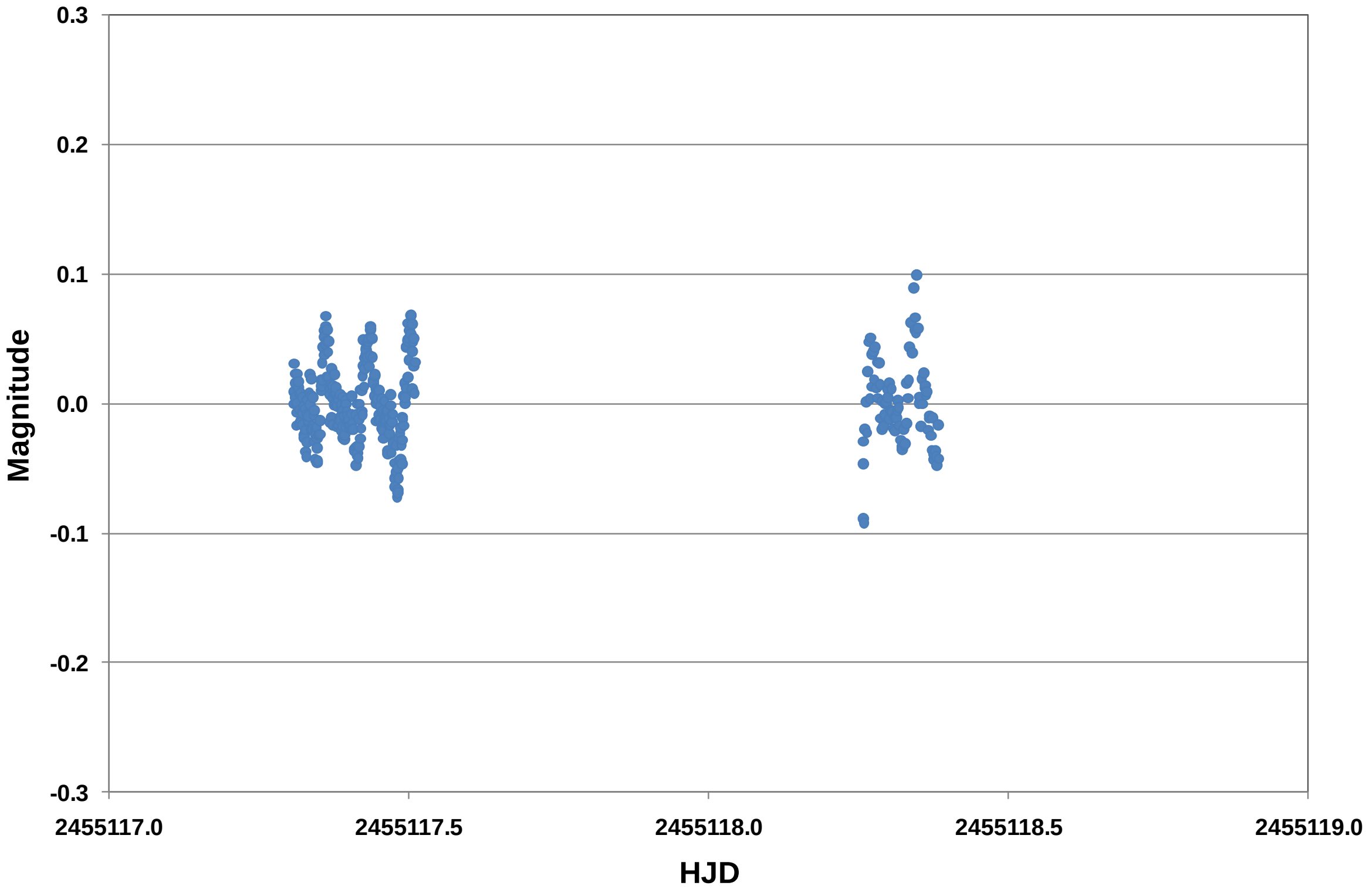

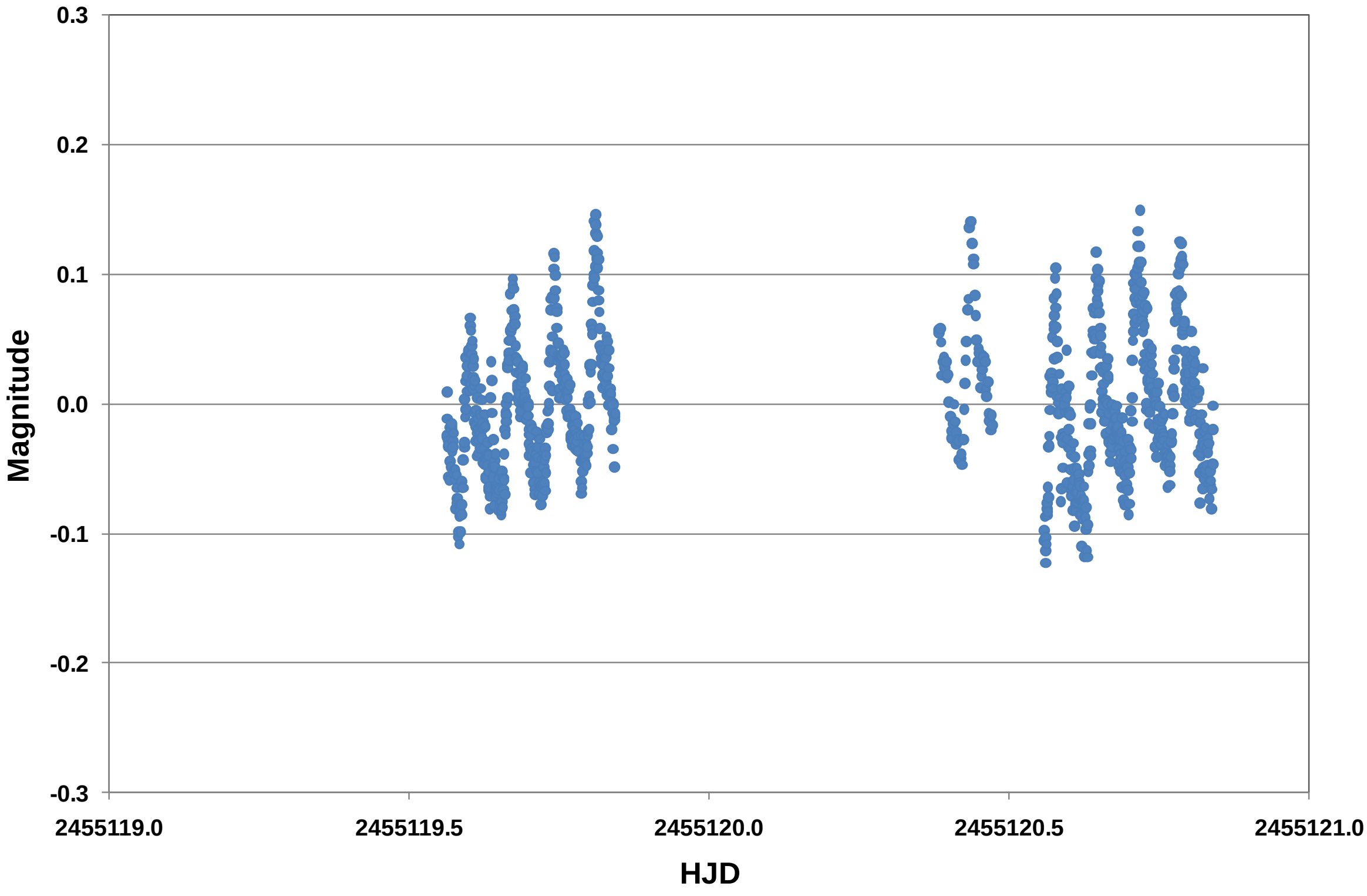

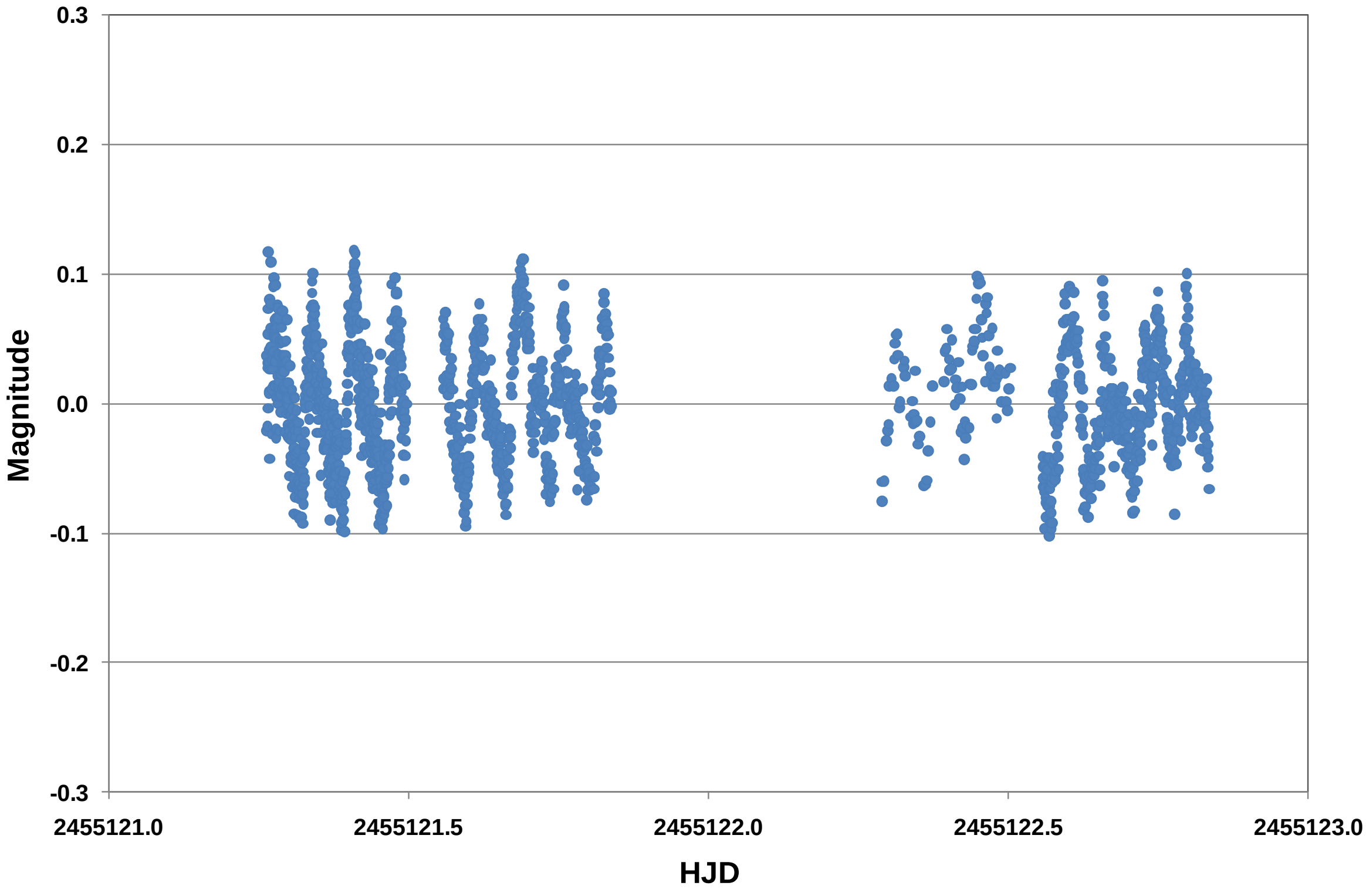

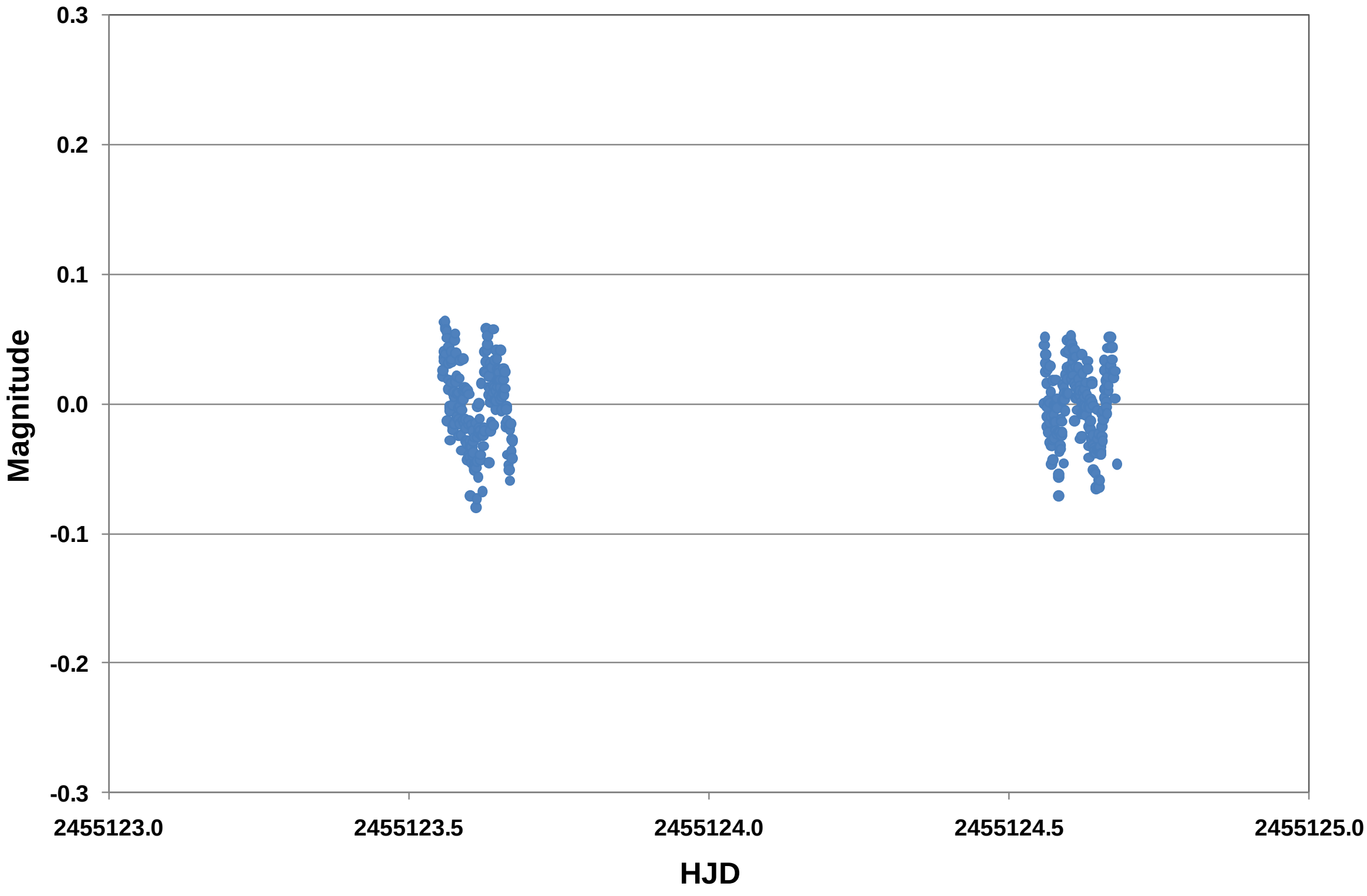

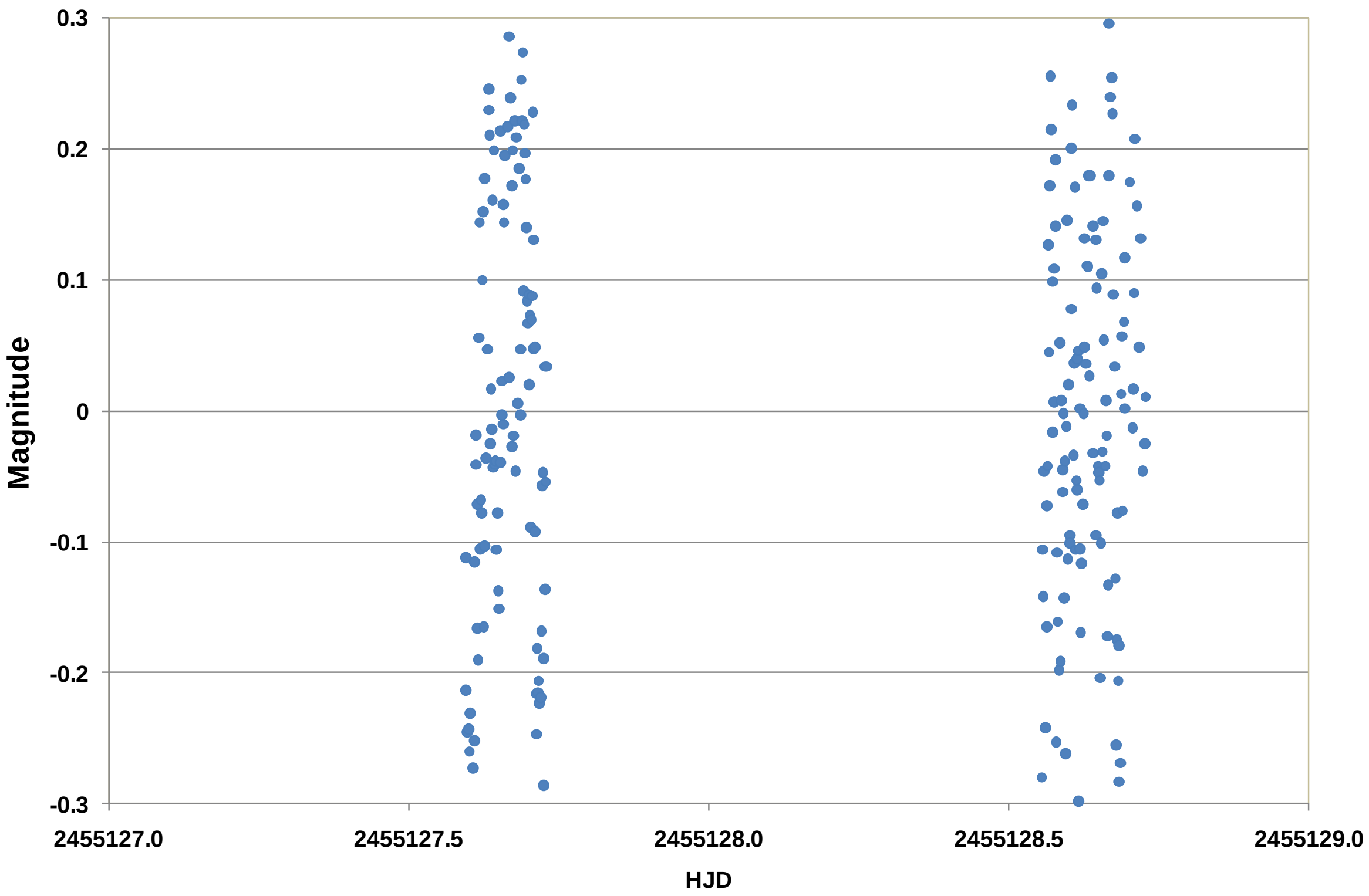

**Figure 3: Time series photometry during the outburst of ASAS J224349+0809.5**

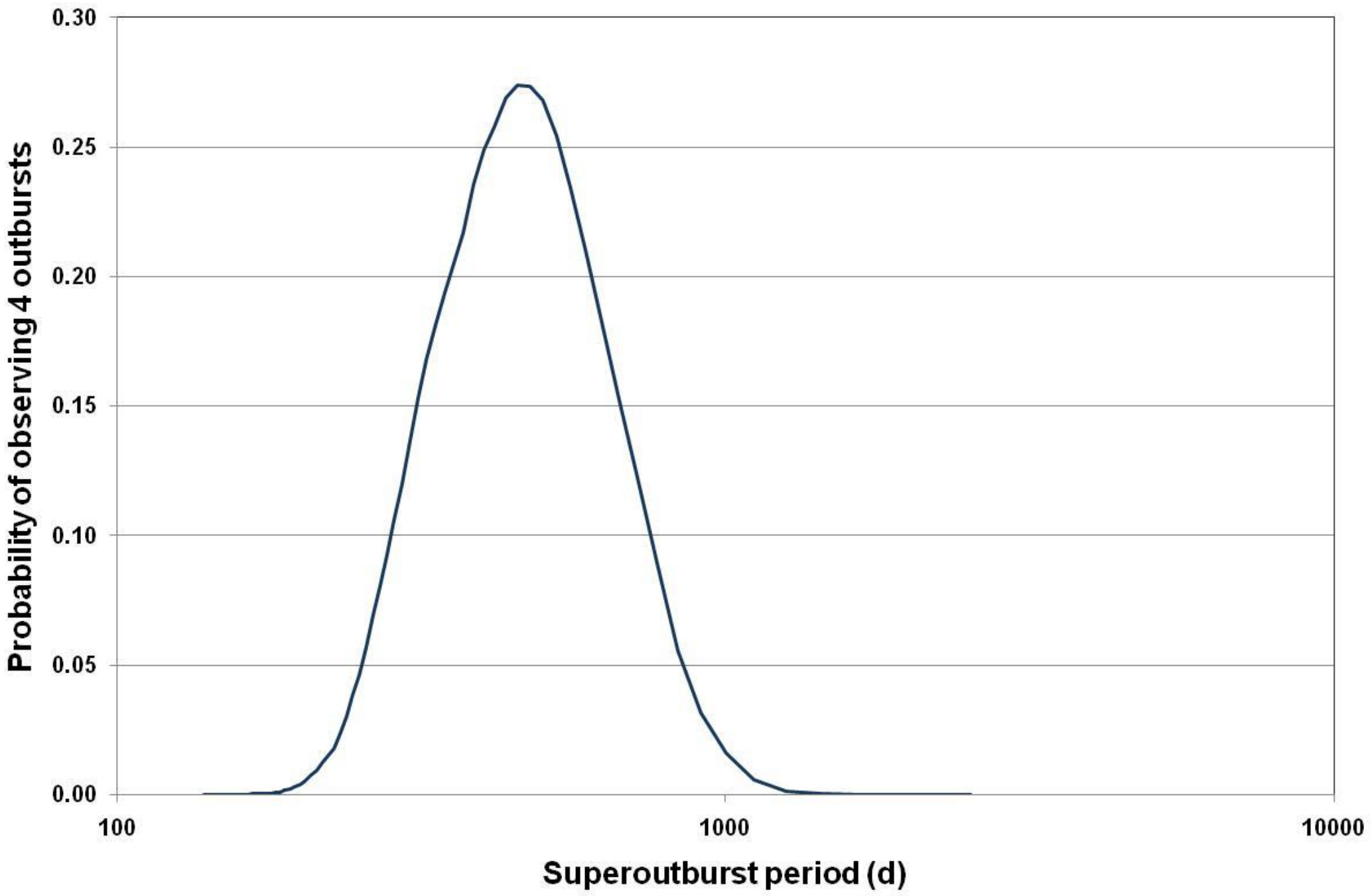

**Figure 4: Probability of observing 4 outbursts for a given length of superoutburst period**